\documentclass{article} %

\usepackage{fullpage}

\usepackage{enumitem}
\setlist{nosep, leftmargin=14pt}
 
\usepackage{mwe} %
\usepackage{bm}
\usepackage{amsfonts}
\usepackage{url}            %
\usepackage{booktabs}       %
\usepackage{amsfonts}       %
\usepackage{nicefrac}       %
\usepackage{microtype}     %
\usepackage{xcolor}         %
\usepackage{bm}         %
\usepackage{graphicx}  		%
\usepackage{amsmath} %
\usepackage{cite} %
\usepackage{diagbox}
\usepackage{enumitem}
\usepackage{siunitx}
\usepackage{multirow}
\usepackage{tabularx}
\usepackage{hhline}
\usepackage{arydshln}		%
\usepackage{xcolor}
\usepackage{mathtools} %
\usepackage{amsthm}			%
\usepackage{amssymb}			%
\usepackage{algorithmic}	%
\usepackage{algorithm}

\def\lim{\mathop{\mathrm{lim}}} %

\newcommand{\norm}[1]{\left\lVert#1\right\rVert}

\def\ebm{{\bm{e}}}

\def\xbm{{\bm{x}}}

\def\ybm{{\bm{y}}}

\def\thetabm{{\bm{\theta}}}

\def\Abm{{\bm{A}}}

\def\Fbm{{\bm{F}}}
\def\Sbm{{\bm{S}}}

\def\Ibm{{\bm{I}}}

\def\Mbm{{\bm{M}}}

\def\E{\mathbb{E}}

\def\Ncal{{\mathcal{N}}}

\title{A Self-supervised Diffusion Bridge for MRI Reconstruction}
\author{Harry Gao$^*$, Weijie Gan$^*$, Yuyang Hu, Hongyu An, and Ulugbek S. Kamilov\\
\small Washington University in St.~Louis, MO, USA\\
$^{\footnotesize *}$\small These authors contributed equally.\\
\small\texttt{\{harrygao, weijie.gan, h.yuyang, hongyuan, kamilov\}@wustl.edu}
}
\date{}

\begin{document}
\maketitle

\begin{abstract}
    Diffusion bridges (DBs) are a class of diffusion models that enable faster sampling by interpolating between two paired image distributions.
    Training traditional DBs for image reconstruction requires high-quality reference images, which limits their applicability to settings where such references are unavailable.
    We propose \emph{SelfDB} as a novel \emph{self-supervised} method for training DBs directly on available noisy measurements without any high-quality reference images. 
    SelfDB formulates the diffusion process by further sub-sampling the available measurements two additional times and training a neural network to reverse the corresponding degradation process by using the available measurements as the training targets. 
    We validate SelfDB on compressed sensing MRI, showing its superior performance compared to the denoising diffusion models.
\end{abstract}

\section{Introduction}
Many problems in biomedical imaging can be viewed as inverse problems, where the goal is to reconstruct an unknown image from its noisy and incomplete measurements. For instance, in compressed sensing MRI (CS-MRI)~\cite{Lustig.etal2007}, the goal is to reconstruct high-quality images from undersampled k-space measurements. Numerous algorithms have been proposed to address inverse problems over the past years with the current research focused on deep learning (DL)-based methods~\cite{Ongie.etal2020,Aggarwal.etal2019}. Among DL approaches, diffusion models have gained popularity for their ability to capture complex data distributions and produce images with impressive perceptual quality~\cite{Daras.etal2024,liu2023dolce}.
  
Denoising diffusion models (DDMs) are a class of methods that define a diffusion process from the ground truth image distribution to the standard Gaussian distribution~\cite{Ho.etal2020}. A deep neural network (DNN) is then trained to reverse this process, allowing generation of images from randomly generated Gaussian noise. Recent studies have shown the potential of DDMs on the inverse problems via posterior sampling~\cite{Chung.etal2023} or conditional generation~\cite{Mao.etal2023a}. However, DDMs are known to be slow during inference due to many inference steps. Diffusion bridges (DBs) are alternatives to DDMs based on a diffusion process that bridges the ground truth image distribution directly to the measurement distribution~\cite{Chung.etal2023d,Delbracio.Milanfar2024}. Once a DNN is trained to reverse the this process, inverse problems can be solved by inferring from the pre-trained DB starting from the unseen measurements. Recent work has shown that DBs can outperform DDMs using much fewer inference steps~\cite{Chung.etal2023d}.

Despite the success of DBs, training them requires high-quality ground-truth images, which may not be available in some applications. On the other hand, recent works on self-supervised learning has investigated training DNNs without any ground-truth images~\cite{Yaman.etal2020,Gan.etal2023,Chen.etal2021}. In particular, a recent paper~\cite{Daras.etal2023a} has proposed a self-supervised method for training traditional DDMs. To the best of our knowledge, self-supervised learning has not been explored in DBs. We address this gap by proposing a new self-supervised method for training diffusion bridges (SelfDB). SelfDB is based on a new diffusion process that further degrades the available measurement two additional times using two different operators. This allows training a DNN to reverse the SelfDB diffusion process using only the available measurements as targets, eliminating the need for ground-truth images. We validate SelfDB on the task of compressed sensing MRI, showing that SelfDB can achieve superior performance compared to a self-supervised DDM.

\section{Background}

We consider an inverse problem for recovering an unknown image $\xbm$ from its noisy and incomplete measurement
\begin{equation}
    \label{equ:inverse}
    \ybm = \Mbm\Abm\xbm + \ebm\ ,
\end{equation}
where $\Abm$ is the measurement matrix, $\ebm$ is the measurement noise, and $\Mbm$ is a binary sub-sampling matrix. 
The matrix $\Abm^\mathsf{H}$ denotes a conjugate transpose that maps measurements back to the image domain.
In CS-MRI, $\Abm$ can be expressed as $\Abm = \Fbm\Sbm$, where $\Fbm$ is the Fourier transform, and $\Sbm$ consists of coil sensitivity maps. Our goal is to address the ill-posed nature of this inverse problem by training a DB using only measurements $\ybm$ without any clean images $\xbm$.

\subsection{Denoising Diffusion Models}

DDMs consider the diffusion process from the image $\xbm$ at $t=0$ to standard Gaussian noise $\bm{\epsilon} \sim \Ncal(0, \Ibm)$ at $t=T$
\begin{equation}
    \xbm_t = \sqrt{\bar{\alpha}_t}\xbm+\sqrt{1-\bar{\alpha}_t}{\bm \epsilon}\ \ \text{for}\ \ t=1,\ ...\ T,
\end{equation}
where $\bar{\alpha}_t = \prod_{i=1}^t\alpha_i$, and $\{\alpha_t\}_{t=1}^T$ are pre-defined parameters.
The goal of DDMs is to train a DNN $\bm{f}_\thetabm$ to reverse this process, enabling it to obtain a clean image from a noise vector $\xbm_T$ sampled from $\Ncal(0, \Ibm)$. Specifically, the DNN is trained to denoise noisy images $\xbm_t$ across all different $t$
\begin{equation}
    \E_{\xbm,\bm{\epsilon},t,\xbm_t} \big[\norm{\bm{f}_\thetabm(\xbm_t, t)-\bm{x}}_2^2\big]\ .
    \label{equ:training-DDM}
\end{equation}
During inference, $\xbm$ is estimated from $\xbm_T$ as~\cite{Ho.etal2020}
\begin{equation}
    \begin{split}
        \xbm_{t-1} &= \mu_{1,t} \xbm_{t} + \mu_{2,t} \bm{f}_\thetabm(\xbm_t, t) + \Sigma_t\bm{\epsilon} \\
    \end{split}
    \label{equ:sampling-DDM}
    \end{equation}
where $\mu_{1,t}=\sqrt{\alpha_t}(1-\bar{\alpha}_{t-1})/(1-\bar{\alpha}_t)$, $\mu_{2,t}=\sqrt{\bar{\alpha}_{t-1}}(1-\alpha_t)/(1-\bar{\alpha}_t)$ and $\Sigma_t=(1-\alpha_t)(1-\bar{\alpha}_{t-1})/(1-\bar{\alpha}_{t})$.
The core concept of this reverse process is to iteratively reduce noise in the image until a clean image is obtained.
This framework is designed for \emph{unconditional} image generation. In inverse problems, one can either integrate a measurement consistency term into the reverse iteration~\eqref{equ:sampling-DDM} for posterior sampling~\cite{Chung.etal2023} or include the measurements as additional inputs to the network during training (denoted as~$\bm{f}_\thetabm(\xbm_t, t, \ybm)$)~\cite{Saharia.etal2022a,Mao.etal2023a}.

A limitation of traditional DDMs is that they rely on ground-truth images as targets for training. To address this, a recent work on \emph{ambient diffusion}\cite{Daras.etal2023a} uses a new operator $\Mbm'\Abm$ to degrade $\xbm_t$ such that the resulting diffusion process ${\ybm_{t}}$ depends only on the measurements $\ybm$
\begin{equation}
    \label{equ:ambient-dm}
    \begin{aligned}
        \ybm_{t} & = \Mbm'\Abm(\sqrt{\bar{\alpha}_t}\xbm+\sqrt{1-\bar{\alpha}_t}{\bm \epsilon})\\
        & =\sqrt{\bar{\alpha}_t}\Mbm'\ybm + \sqrt{1-\bar{\alpha}_t}\Mbm'\Abm{\bm \epsilon}\ ,
    \end{aligned}
\end{equation}
where $\Mbm'$ is a further sub-sampling matrix of $\Mbm$. A DNN is then trained to reconstruct $\xbm$ from $\ybm_t$ for each $t$ using the measurements as training targets
\begin{equation}
    \E_{\ybm,\bm{\epsilon},t,\ybm_{t}} \big[\norm{\Mbm\Abm\bm{f}_\thetabm(\ybm_{t}, t)-\bm{y}}_2^2\big]\ .
    \label{equ:training-ambient}
\end{equation}
The effectiveness of this training loss has been demonstrated in~\cite{Daras.etal2023a} as well as other related recent work on self-supervised learning for the inverse problems~\cite{Yaman.etal2020,Gan.etal2023}.
At inference time, the iteration in~\eqref{equ:sampling-DDM} is applied by first degrading $\xbm_t$ using $\Mbm'\Abm$ and then feeding it to the DNN.

\subsection{Diffusion Bridges}
We consider DBs that formulate the diffusion process from the ground truth image $\xbm$ at $t=1$ to the measurement $\ybm$ at $t=0$ through a simple linear combination~\cite{Delbracio.Milanfar2024}
\begin{equation}
    \xbm_t = (1-t) \xbm+ t\Abm^\mathsf{H}\ybm + \sigma_t{\bm \epsilon}\ \ \text{for}\ \ t\in[0,1]\ ,
\end{equation}
where ${\sigma_t}$ is a standard deviation. Like DDMs, DBs aim to train a neural network to reverse this process, allowing iterative image reconstruction from the measurement $\ybm$ (i.e., $\xbm_t$ at $t = 1$). Because the reverse process of DBs originates from the measurement domain---a closer approximation to the ground truth domain than the standard Gaussian domain in DDM---it offers superior performance with less inference steps~\cite{Chung.etal2023d}. The supervised training loss for DBs can be formulated as 
\begin{equation}
    \E_{\xbm,\bm{\epsilon},t,\xbm_t} \big[\norm{\bm{f}_\thetabm(\xbm_t, t)-\bm{\xbm}}_2^2\big]\ .
\end{equation}
During inference, with $\xbm_1 = \Abm^\mathsf{H}\ybm + \sigma_1\bm{\epsilon}$, the DB iteration is given by $\E[\xbm_{t-\delta}|\xbm_t]$ (see also Proposition 4.1 in \cite{Delbracio.Milanfar2024})
\begin{equation}
    \xbm_{t-\delta} = \frac{\delta}{t} \bm{f_\theta}(\xbm_t, t) + (1-\frac{\delta}{t})\xbm_t+(t-\delta)\sqrt{\sigma^2_{t-\delta}-\sigma_{t}^2}\bm{\epsilon}\ ,
\end{equation}
where $\delta$ is the iteration step size. The key concept behind this reverse process is that it gradually improves the image in small steps, making the reconstruction task less and less challenging compared to directly solving the inverse problem.

\begin{algorithm}[t]
\caption{Inference in SelfDB}
\label{alg:selfdb}
\begin{algorithmic}[1]
\STATE \textbf{input}: $\ybm$, $\bm{f_\theta}$, $\delta$, $\{\sigma_t\}$.
\STATE $\ybm_1 = \Mbm'\ybm + \sigma_1\bm{\epsilon}$
\FOR{$t = 1$ to $0$ with step $\delta$}
\STATE $\bm{\epsilon}\sim\Ncal(0,\Ibm)$
\STATE $\hat{\xbm} = \bm{f_\theta}(\ybm_t, t)$
\IF{$t>0$}
\STATE $\hat{\ybm} = \overline{\Mbm}\Abm\hat{\xbm}$ 
\ELSE
\RETURN $\hat{\xbm}$
\ENDIF
\STATE $\ybm_{t-\delta} = \frac{\delta}{t} \hat{\ybm} + (1-\frac{\delta}{t})\ybm_t+(t-\delta)\sqrt{\sigma^2_{t-\delta}-\sigma_{t}^2}\bm{\epsilon}\ .$
\ENDFOR
\end{algorithmic}
\end{algorithm} 

\begin{figure*}
    \centering
    \includegraphics[width=.92\linewidth]{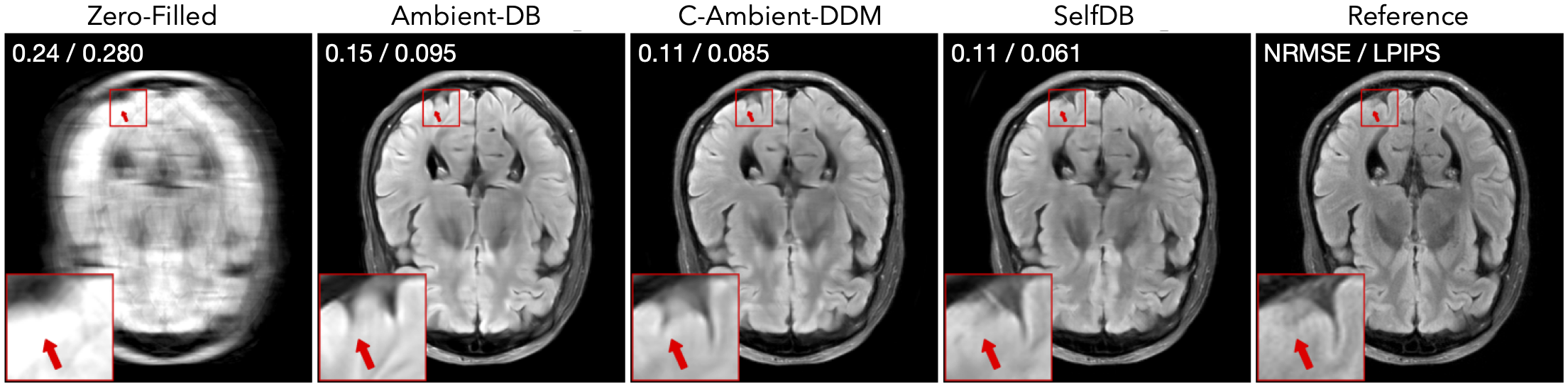}
    \caption{Visual comparisons of \emph{Ambient-DB}, conditional ambient diffusion (\emph{C-Ambient-DDM}), and the proposed SelfDB. \emph{C-Ambient-DDM} is an existing self-supervised approach for denoising diffusion models, while \emph{Ambient-DB} is its direct extension into the DB framework. NRMSE and LPIPS values for each method are labeled in the top-left corner of the images. Note how SelfDB yields images with details closely matching the reference, highlighted by the read arrow in the zoomed-in region.}
    \label{fig:visual-results}
\end{figure*}

\section{Method}
We now introduce our proposed self-supervised algorithm for DBs. We begin by showing why a direct application of ambient diffusion to DBs, which we call \emph{Ambient-DB}, is suboptimal in CS-MRI. Then, we present our proposed SelfDB approach.

The core concept behind ambient diffusion is to construct a new diffusion process that eliminates the need for ground truth data during training as in~\eqref{equ:ambient-dm}. A direct adaptation of this concept into DBs is to apply the same operator $\Mbm'\Abm$ to $\xbm_t$
\begin{equation}
    \label{equ:native}
    \begin{aligned}
        \ybm_{t} & = \Mbm'\Abm\xbm_t = \Mbm'\Abm((1-t) \xbm+ t\Abm^\mathsf{H}\ybm + \sigma_t{\bm \epsilon}) \\
        & = \Mbm'\ybm + \sigma_t\Mbm'\Abm\bm{\epsilon}\ ,
    \end{aligned}
    \end{equation}
where the last line holds because $\Mbm'$ is a further sub-sampling matrix of $\Mbm$ and $\Abm$ is orthogonal in CS-MRI. Since only a small amount of noise is added during training~\cite{Delbracio.Milanfar2024}, $\ybm_{t}$ does not establish a diffusion process that transitions between distinct data, as intended by the concept of DBs. Instead, different values of $t$ correspond to nearly identical sub-sampled data $\Mbm'\ybm$, rendering the reverse process effectively as repeated solutions of the same inverse problem. Our experimental results in Section~\ref{sec:results} also show that this diffusion process leads to suboptimal performance.

To address this, the proposed SelfDB approach considers a new sub-sampling operator $\overline{\Mbm}$ that provides an intermediate level of subsampling between $\Mbm$ and $\Mbm'$. This enables a diffusion process for SelfDB that transitions from intermediately sub-sampled data at $t=0$ to highly sub-sampled data at $t=1$ 
\begin{equation}
    \ybm_t^{} = (1-t) \overline{\Mbm}{\ybm} + t\Mbm'\ybm_{} + \sigma_t{\bm \epsilon}\ .
\end{equation}
This modified diffusion process allows for a reverse process that does not directly reconstruct $\Mbm'\ybm_{}$. Instead, it incrementally incorporates information from $\overline{\Mbm}\ybm_{}$, making the inverse problem progressively less challenging. In alignment with~\eqref{equ:training-ambient}, we define our training loss to map $\ybm_t^{}$ to the measurement $\ybm$ as
\begin{equation}
    \E_{\ybm,\bm{\epsilon}, t, \ybm_{t}^{}} \big[\norm{\Mbm\Abm\bm{f}_\thetabm(\ybm_t^{}, t)-\bm{\ybm}}_2^2\big]\ .
\end{equation}
The inference of SelfDB is summarized in Algorithm~\ref{alg:selfdb}, where the main update iteration (line 11) is given by $\E[\ybm_{t-\delta}|\ybm_t]$.

\begin{table}[t]
    \caption{Average NRMSE, SSIM, and LPIPS values for SelfDB, \emph{Ambient-DB}, and \emph{C-Ambient-DDM} on the test set. This table demonstrates that SelfDB outperforms these baselines across both distortion and perceptual metrics.}
    \vspace{5pt}
    \centering
    \small
    \renewcommand\arraystretch{1.1}
    \setlength{\tabcolsep}{5pt}
    \begin{tabular}{ccccc}
    \toprule
    {Metrics}                   & NRMSE $\downarrow$ & SSIM $\uparrow$ & LPIPS $\downarrow$    \\
    \cmidrule{1-5}
    \emph{Ambient-DB} & 0.172 & 0.782 & 0.069 \\
    \emph{C-Ambient-DDM} & 0.113 & 0.833 & 0.056 \\
    \emph{SelfDB}    & \textbf{0.108} & \textbf{0.882} & \textbf{0.036} \\
    \bottomrule 
    \end{tabular}
    \label{tab:visual-results}
\end{table}

\section{Numerical Validation}
\label{sec:results}
We validated SelfDB in CS-MRI using fully-sampled T2-weighted MR brain data from around 1000 subjects in the fastMRI dataset~\cite{Knoll.etal2020a}. These subjects were split into 800 for training, 100 for validation, and 100 for testing. For each subject, we used the middle 7 to 9 slices in the transverse plane. The coil sensitivity map $\Sbm$ was estimated via ESPIRiT~\cite{Uecker.etal2014}. 
For our experiments, we simulated Cartesian sampling operators with sampling rates of $25\%$, $16.6\%$, and $12.5\%$ as $\Mbm$, $\overline{\Mbm}$, and $\Mbm'$, respectively. Specifically, we trained SelfDB on a dataset with $25\%$ sub-sampled MRI data and tested it on data sub-sampled at $12.5\%$. During testing, we considered a small inference step of 4. We evaluated the results using \emph{normalize-root-mean-square-error(NRMSE)}/\emph{structural similarity index measure (SSIM)} and \emph{learned perceptual image patch similarity (LPIPS)} as distortion based and perceptual metrics, respectively.

We compare SelfDB against \emph{Ambient-DB} and ambient diffusion~\cite{Daras.etal2023a}. Each baseline model was trained using the same experimental setup. Notably, ambient diffusion was originally designed for unconditional image generation. To adapt it for inverse problems, we implemented it within a conditional diffusion model framework---termed \emph{C-Ambient-DDM}---by incorporating measurements as additional inputs to the DNN.

\begin{figure}
    \centering
    \includegraphics[width=.73\linewidth]{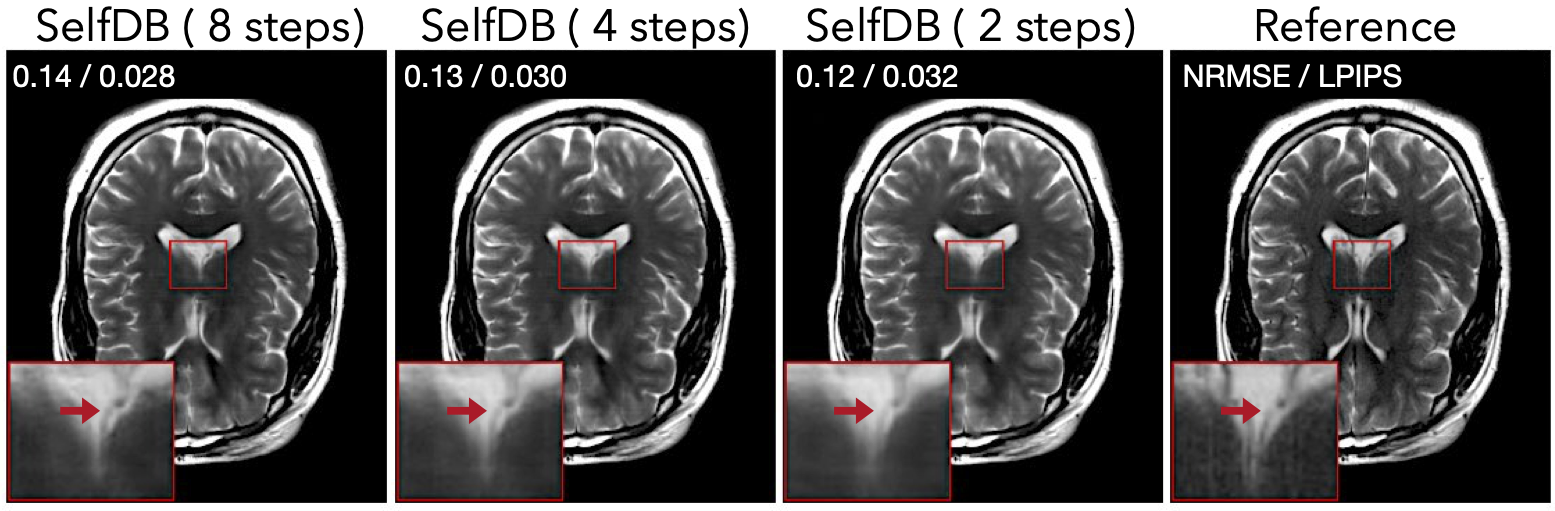}
    \caption{Visual results of SelfDB with different inference steps. Best viewed in digital format. This figure demonstrates that SelfDB empirically provides a perception-distortion trade-off. Notably, perceptual quality (LPIPS) improves as the number of inference steps increases, while distortion with respect to the reference (NRMSE) also rises.}
    \label{fig:visual-tradeoff}
\end{figure}

Figure~\ref{fig:visual-results} presents visual results comparing SelfDB with baseline methods, showing that SelfDB can reconstruct images with both high perceptual quality and fine details closely matching the reference. Table~\ref{tab:visual-results} summarizes the test metrics for SelfDB and the baseline methods, demonstrating that SelfDB can gain superior performance in both distortion-based and perceptual metrics. In addition to its promising performance, SelfDB also empirically provides a perception-distortion trade-off that can be adjusted by varying the number of inference steps. As illustrated in Figure~\ref{fig:visual-tradeoff} and shown in Table~\ref{tab:trade-off}, increasing the sampling steps improves perceptual quality while introducing more distortion. This trade-off is also discussed in other DB-based methods~\cite{Delbracio.Milanfar2024}. The ability of SelfDB to balance the perception-distortion trade-off may be particularly valuable in applications where either quantitative accuracy or perceptual quality is preferred.

\begin{table}[t]
    \caption{Average NRMSE, SSIM, and LPIPS values for SelfDB on the test set across different inference steps. This table shows that the number of inference step in SelfDB empirically controls the perception-distortion trade-off.}
    \vspace{5pt}
    \centering
    \small
    \renewcommand\arraystretch{1.1}
    \setlength{\tabcolsep}{5pt}
    \begin{tabular}{ccccc}
    \toprule
    {Metrics}                   & NRMSE $\downarrow$ & SSIM $\uparrow$ & LPIPS $\downarrow$    \\
    \cmidrule{1-5}
    \emph{SelfDB} (2 steps)    & \textbf{0.101} & \textbf{0.888} & 0.039 \\
    \emph{SelfDB} (4 steps)    & 0.108 & 0.882 & 0.036 \\
    \emph{SelfDB} (8 steps)    & 0.118 & 0.875 & \textbf{0.035} \\
    \bottomrule 
    \end{tabular}
    \label{tab:trade-off}
\end{table}

\section{Conclusion}
This paper presents SelfDB as the first self-supervised learning algorithm for diffusion bridges. SelfDB proposes a novel diffusion process from intermediately to severely subsampled data. This data is generated by applying two different subsampling operators to the available measurements. The core innovation in this approach lies in \emph{(a)} a diffusion process that bypasses the need for ground truth images and \emph{(b)} allowing the training of a DNN to reverse this process using raw measurements as training targets. We validated SelfDB in CS-MRI. The experimental results show that SelfDB can gain superior performance than its denoising diffusion counterpart under small inference steps. 

\section{Compliance with Ethical Standards}
This research study was conducted retrospectively using human subject data made available in open access by~\cite{Knoll.etal2020a}. Ethical approval was not required as confirmed by the license attached with the open access data.

\section{Acknowledgments}
This work was supported by the NSF CAREER award under CCF-2043134. This work is also supported by NIH R01 EB032713, RF1NS116565, R21NS127425 and NIH R01HL129241.

{

}

\end{document}